\documentclass[aps,twocolumn,notitlepage,superscriptaddress]{revtex4-1}

\usepackage{amsfonts,amsmath,amssymb,amsthm,dsfont,float,graphicx,hyperref,hyphenat,lmodern,mathtools,microtype,multirow,pgfplots,slashbox,xcolor}
\usepackage[UKenglish]{babel}
\usepackage[capitalize,nameinlink]{cleveref}
\usepackage[T1]{fontenc}
\usepackage[utf8]{inputenc}
\usepackage[UKenglish]{isodate}

\definecolor{darkred}{rgb}{0.5, 0, 0}
\definecolor{darkgreen}{rgb}{0, 0.5, 0}
\definecolor{darkblue}{rgb}{0, 0, 0.5}
\hypersetup{colorlinks,breaklinks,linkcolor=darkred,citecolor=darkgreen,urlcolor=darkblue}

\DeclareMathOperator{\Tr}{Tr}
\DeclareMathOperator{\conv}{conv}

\renewcommand{\leq}{\leqslant}
\renewcommand{\geq}{\geqslant}

\newcommand{\bd}{\mathbf{d}}
\newcommand{\be}{\mathbf{e}}
\newcommand{\bbf}{\mathbf{f}}
\newcommand{\bp}{\mathbf{p}}
\newcommand{\br}{\mathbf{r}}
\newcommand{\bx}{\mathbf{x}}
\newcommand{\bv}{\mathbf{v}}
\newcommand{\GHZ}{\mathrm{GHZ}}
\newcommand{\id}{\mathds{1}}
\newcommand{\ket}[1]{|#1\rangle}
\newcommand{\ketbra}[1]{|#1\rangle \langle#1|}
\newcommand{\me}{\mathrm{e}}
\newcommand{\mi}{\mathrm{i}}
\newcommand{\mL}{\mathcal{L}}
\newcommand{\va}{\vec{a}}
\newcommand{\vb}{\vec{b}}
\newcommand{\vx}{\vec{x}}

\begin{document}

\title{Symmetric multipartite Bell inequalities via Frank-Wolfe algorithms}
\author{Sébastien Designolle}
\affiliation{Zuse-Institut Berlin, Takustraße 7, 14195 Berlin, Germany}
\author{Tamás Vértesi}
\affiliation{\mbox{MTA ATOMKI Lend\"ulet Quantum Correlations Research Group, HUN-REN Institute for Nuclear Research, Debrecen, 4001, Hungary}}
\author{Sebastian Pokutta}
\affiliation{Zuse-Institut Berlin, Takustraße 7, 14195 Berlin, Germany}
\date{5th October 2026}

\begin{abstract}
  In multipartite Bell scenarios, we study the nonlocality robustness of the Greenberger-Horne-Zeilinger (GHZ) state.
  When each party performs planar measurements forming a regular polygon, we exploit the symmetry of the resulting correlation tensor to drastically accelerate the computation of (i) a Bell inequality via Frank-Wolfe algorithms, and (ii) the corresponding local bound.
  The Bell inequalities obtained are facets of the symmetrised local polytope and they give the best known upper bounds on the nonlocality robustness of the GHZ state for three to ten parties.
  Moreover, for four measurements per party, we generalise our facets and hence show, for any number of parties, an improvement on Mermin's inequality in terms of noise robustness.
  We also compute the detection efficiency of our inequalities and show that some give rise to activation of nonlocality in star networks, a property that was only shown with an infinite number of measurements.
\end{abstract}

\maketitle

\section{Introduction}

Bell nonlocality of multipartite quantum states is one of the most counter-intuitive aspects of nature~\cite{Bel64}.
Measurements on entangled quantum states are indeed able to produce correlations that cannot be explained by any local classical mechanism.
These nonlocal correlations can be witnessed in experiments by violating Bell inequalities~\cite{Bel64,CHSH69}.
In addition to its fundamental importance, Bell nonlocality also enables a technology based on the so-called device-independent paradigm of quantum information processing~\cite{BCP+14,Sca19}.
However, in order to certify the nonlocal nature of multipartite correlations and make it useful in real-life applications, one quickly runs into the problem of exponential scaling of the parameter space with the number of parties.
Already for Bell inequalities with binary settings and outcomes, the number of extremal vertices and the dimension of the Bell polytope grow exponentially with the number of parties.
In order to tackle this problem, we use two tools: a very efficient algorithm that solves constrained quadratic optimisation by only resorting to linear optimisations, and the exploitation of symmetries present in multipartite quantum states and measurements.
These two different tools, as we will see, significantly reduce the computational complexity of the problem and allow us to go well beyond the standard scenarios tractable using linear programming~\cite{GLZ+10,GGH+14,PBM+22}, or even the results obtained with Frank-Wolfe algorithms~\cite{BNV16,DIB+23}.

More precisely, we show how symmetries can be exploited to accelerate both the Frank-Wolfe approach from~\cite{DIB+23} and the computation of the local bounds of the resulting Bell inequalities.
The symmetries considered are not ambient (they do not depend on the scenario itself) but embedded in the specific instance we consider, namely, the multipartite GHZ state where each party performs measurements forming a regular polygon in the XY plane of the Bloch sphere.
We obtain facets of the symmetrised local polytope for up to $N=10$ parties and $m=9$ measurements and give some consequences of these results.
For instance, the case just mentioned allows us to show the activation of nonlocality in a star network, a fact that was known only by using an infinite number of measurements~\cite{SSB+05}.
Moreover, the inequalities obtained with our method for $m=4$ appear to follow a pattern that we can generalise to all $N$, improving on the asymptotic results given by Mermin's inequality~\cite{Mer90}.

After having properly defined the notions involved in \cref{sec:preliminaries}, we illustrate the main symmetry reductions with a small example in \cref{sec:example} before generalising the arguments to arbitrary scenarios in \cref{sec:group,sec:polytope,sec:fw,sec:bound}.
Then we give the main results in \cref{sec:results} before discussing some consequences in \cref{sec:consequences}.

\section{Preliminaries}
\label{sec:preliminaries}

We consider a scenario in which $N$ parties, labelled by $n\in[N]$, upon receiving inputs $x_n\in[m]$, give answers $a_n=\pm1$ according to a strategy that they predefined.
When repeating this for many rounds, one can construct the probabilities $p(a_1\ldots a_N|x_1\ldots x_N)$ corresponding to this strategy.
In this work, marginals will always be zero and we work in the correlation notation~\cite{Sca19}.
If the parties have access to some shared quantum state, they may obtain correlations that cannot be explained with local means.
Geometrically, in this finite scenario, this amounts to saying that these correlations are outside of the local polytope defined as the convex hull of local deterministic strategies.

Formally, for a choice of local deterministic strategies $\va^{(n)}\coloneqq (a^{(n)}_1\ldots a^{(n)}_m)\in\{\pm1\}^m$ for all $n\in[N]$, the deterministic strategy $\bd^{\va^{(1)}\ldots \va^{(N)}}$ has elements
\begin{equation}
  d^{\va^{(1)}\ldots \va^{(N)}}_{x_1\ldots x_N}\coloneqq\prod_{n=1}^Na^{(n)}_{x_n},
  \label{eqn:deterministic}
\end{equation}
and the local polytope is defined as
\begin{equation}
  \mL^{(m)}_N\coloneqq\conv\left\{\bd^{\va^{(1)}\ldots \va^{(N)}}~\Big|~\va^{(1)}\ldots \va^{(N)}\in\{\pm1\}^m\right\}.
  \label{eqn:mL}
\end{equation}
A point inside this polytope hence admits a convex decomposition in terms of these deterministic strategies.
Such a decomposition is called a local model.
Conversely, points outside of the polytope may be detected by separating hyperplanes, named Bell inequalities.

Consider the following $N$-partite GHZ state~\cite{GHZ89}
\begin{equation}
  \ket{\GHZ_N}\coloneqq\frac{\ket{0\ldots0}+\ket{1\ldots1}}{\sqrt2},
  \label{eqn:GHZ}
\end{equation}
where $\ket{0}$ and $\ket{1}$ correspond to eigenvectors of $\sigma_Z$; in other words, they are oriented vertically in the Bloch sphere.
We are primarily interested in the \emph{nonlocality robustness} of this state~\cite{PMS16}.
In general, for a density matrix $\rho$, this quantity is defined to be
\begin{equation}
  v_c^\rho\coloneqq\inf\left\{v~\Big|~\exists m,\exists \vec{A},v\bp^{\rho,\vec{A}}\not\in\mL^{(m)}_N\right\},
  \label{eqn:vc}
\end{equation}
where, given the observables $\vec{A}\coloneqq\{\{A_{x_n}^{(n)}\}_{x_n}\}_n$, the correlation tensor $\bp^{\rho,\vec{A}}$ has elements
\begin{equation}
  p^{\rho,\vec{A}}_{x_A\ldots x_N}\coloneqq\Tr \left[\left(A_{x_1}^{(1)}\otimes\ldots\otimes A_{x_N}^{(N)}\right)\rho\right],
  \label{eqn:pA}
\end{equation}
according to Born's rule.

In this work, we will detect the nonlocality of the quantum state in \cref{eqn:GHZ} (so $\rho=\ketbra{\GHZ_N}$) by having each party perform $m$ measurements forming a regular polygon in the XY plane of the Bloch sphere~\cite{BNV16}, that is,
\begin{equation}
  A^{(n)}_{x_n}\coloneqq\cos\left(\frac{\pi}{m}x_n\right)\sigma_X+\sin\left(\frac{\pi}{m}x_n\right)\sigma_Y.
  \label{eqn:polygon}
\end{equation}
When doing so, it follows from \cref{eqn:pA} that the resulting correlation tensor, denoted $\br^{(m)}_N$, has elements
\begin{equation}
  r^{(m)}_{x_1\ldots x_N}\coloneqq\cos\left(\frac{\pi}{m}\sum_{n=1}^Nx_n\right).
  \label{eqn:GHZtensor}
\end{equation}

The main focus of this article is to compute the critical visibility
\begin{equation}
  v_m^{\GHZ_N} \coloneqq \max\left\{v~\Big|~v\br^{(m)}_N\in\mL^{(m)}_N\right\},
  \label{eqn:vm}
\end{equation}
which is a well-defined quantity as $\mathbf{0}\in\mL^{(m)}_N$ and $\mL^{(m)}_N$ is compact.
By definition of the nonlocality robustness in \cref{eqn:vc}, one immediately has, for all $m$,
\begin{equation}
  v_c^{\GHZ_N}\leq v_m^{\GHZ_N}.
  \label{eqn:vcup}
\end{equation}

\section{Simple example}
\label{sec:example}

We start with an illustration of the main symmetry reduction techniques when $N=2$ and $m=3$.
In this case, the correlation tensor from \cref{eqn:GHZtensor} simply reads
\begin{equation}
  \br^{(3)}_2=
  \begin{pmatrix}
    1        & \frac12  & -\frac12 \\
    \frac12  & -\frac12 & -1       \\
    -\frac12 & -1       & -\frac12
  \end{pmatrix}.
  \label{eqn:GHZmatrix}
\end{equation}

The first step is to identify the symmetries of this matrix.
They can be seen in \cref{eqn:GHZtensor} and boil down, in this simple case, to the following equalities:
\begin{equation}
  A\br^{(3)}_2 A=\br^{(3)}_2\quad\text{and}\quad B\br^{(3)}_2 B = \br^{(3)}_2,
  \label{eqn:symmetry_example}
\end{equation}
where $A$ corresponds to a cycling with a sign flip of the overflowing element and $B$ acts as a mirror on the last two elements on top of flipping their signs, namely,
\begin{equation}
  A\coloneqq
  \begin{pmatrix}
    0  & 1 & 0 \\
    0  & 0 & 1 \\
    -1 & 0 & 0
  \end{pmatrix}
  \quad\text{and}\quad
  B\coloneqq
  \begin{pmatrix}
    1 & 0  & 0 \\
    0 & 0  & -1 \\
    0 & -1 & 0
  \end{pmatrix}.
\end{equation}
The Reynolds operator $\Gamma$ associated to the action generated by $A$ and $B$ reads
\begin{equation}\nonumber
  \resizebox{\linewidth}{!}{$\Gamma(\bp)\coloneqq\frac16\left(\bp+A\bp A+A^2\bp A^2+B\bp B+BA\bp AB+AB\bp BA\right)$}
\end{equation}
and projects any $3\times 3$ matrix $\bp$ onto the subspace of matrices of the form
\begin{equation}
  \begin{pmatrix}
    \alpha & \beta   & -\beta\\
    \beta  & -\beta  & -\alpha\\
    -\beta & -\alpha & -\beta
  \end{pmatrix}
  \eqqcolon
  \begin{bmatrix} \alpha \\ \beta \end{bmatrix}.
  \label{eqn:subspace_example}
\end{equation}

\begin{figure}[h]
  \centering
  \includegraphics{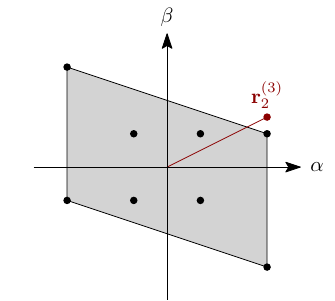}
  \caption{
    The symmetrised local polytope for $N=2$ and $m=3$, that is, its projection on the subspace of matrices of the form of \cref{eqn:subspace_example}.
    The point $\br^{(3)}_2$ outside of the polytope is given in \cref{eqn:GHZmatrix}.
  }
  \label{fig:cross23}
\end{figure}

Now consider the local polytope projected onto this subspace; it is the convex hull of the following eight points:
\begin{equation}
  \begin{bmatrix} -1 \\ -\frac13 \end{bmatrix},
  \begin{bmatrix} -1 \\ 1 \end{bmatrix},
  \begin{bmatrix} 1 \\ -1 \end{bmatrix},
  \begin{bmatrix} 1 \\ \frac13 \end{bmatrix},
  \label{eqn:vertices_extreme_example}
\end{equation}
\begin{equation}
  \begin{bmatrix} -\frac13 \\ -\frac13 \end{bmatrix},
  \begin{bmatrix} -\frac13 \\ \frac13 \end{bmatrix},
  \begin{bmatrix} \frac13 \\ -\frac13 \end{bmatrix},
  \begin{bmatrix} \frac13 \\ \frac13 \end{bmatrix},
  \label{eqn:vertices_example}
\end{equation}
and is depicted in \cref{fig:cross23}.
As can be seen there, only the four points of \cref{eqn:vertices_extreme_example} are extreme points of the projected polytope.
This symmetrised polytope has four facets given by the following inequalities:
\begin{equation}
  \left\langle\pm\!\begin{bmatrix} 1 \\ 0 \end{bmatrix},
  \begin{bmatrix} \alpha \\ \beta \end{bmatrix}\right\rangle\leq3
  \quad\text{and}\quad
  \left\langle\pm\!\begin{bmatrix} 2 \\ 3 \end{bmatrix},
  \begin{bmatrix} \alpha \\ \beta \end{bmatrix}\right\rangle\leq12.
  \label{eqn:facets_example}
\end{equation}
Notice that the right-hand sides in \cref{eqn:facets_example} correspond to the scalar product in the initial nonsymmetrised space, not to the one represented in \cref{fig:cross23}.

The noise robustness is the scalar $v$ at which the point $v\br^{(3)}_2$ enters the local polytope.
In this case it is exactly $\frac45$ as shown by the following equality:
\begin{equation}
  \frac45\br^{(3)}_2=\frac45
  \begin{bmatrix} 1 \\ \frac12 \end{bmatrix}
  =\frac{9}{10}
  \begin{bmatrix} 1 \\ \frac13 \end{bmatrix}
  +\frac{1}{10}
  \begin{bmatrix} -1 \\ 1 \end{bmatrix}.
  \label{eqn:fw_example}
\end{equation}
The decomposition indeed lies precisely on the facet optimally detecting $\br^{(3)}_2$, that is, $\bbf\coloneqq[2,3]$.

When it comes to the computation of the local bound of this Bell inequality, namely, 12 as given in \cref{eqn:facets_example}, the symmetry can also play a role.
Formally, this bound $L_2^{(3)}$ is defined by
\begin{equation}
  L_2^{(3)}\coloneqq\max_{\bd^{\va,\vb}}\left\langle\bbf,\bd^{\va,\vb}\right\rangle=\max_{\va,\vb\in\{\pm1\}^m}\sum_{x,y\in[m]}f_{xy}a_xb_y.
  \label{eqn:L_example}
\end{equation}
Because $\bbf$ is symmetric, i.e., it satisfies $\bbf=\Gamma(\bbf)$, we have
\begin{equation}
  \left\langle\bbf,\bd^{\va,\vb}\right\rangle=
  \left\langle\Gamma(\bbf),\bd^{\va,\vb}\right\rangle=
  \left\langle\bbf,\Gamma(\bd^{\va,\vb})\right\rangle
  \label{eqn:f_sym}
\end{equation}
since $\Gamma$ is self-adjoint.
This ensures that we can restrict the search to the symmetrised deterministic strategies.
Actually, $\vb$ can be fixed once the strategy $\va$ is chosen.
This simplifies the computation of $L_2^{(3)}$ to
\begin{equation}
  L_2^{(3)}=\max_{\va\in\{\pm1\}^m}\sum_{y\in[m]}\Bigg|\sum_{x\in[m]}f_{xy}a_x\Bigg|,
  \label{eqn:L_abs}
\end{equation}
which reduces the number of strategies to the number of orbits of a suitable action discussed below in the general case.
For the simple example of this section, instead of running over the eight possible strategies $\va$, it suffices to consider the following two strategies:
\begin{equation}
  \va=(-1, -1, -1)\quad\text{and}\quad\va=(-1, 1, -1),
  \label{eqn:orbits_example}
\end{equation}
giving rise, respectively, to
\begin{equation}
  \vb=(-1, 1, 1)\quad\text{and}\quad\vb=(1, -1, 1),
  \label{eqn:b_example}
\end{equation}
which both attain the optimal value of 12 here.

\cref{sec:group,sec:polytope,sec:fw,sec:bound} generalise these symmetry arguments to any number of parties $N$ and any number of measurements $m$, showing, in particular, the following reductions:
\begin{enumerate}
  \item[(a)] from $m^N$ to $\lceil\tfrac{m}{2}\rceil$ for the dimension of the space,
  \item[(b)] from $2^{(N-1)(m-1)}$ to $\binom{u_m+N-2}{N-1}$ for the number of strategies to enumerate, where the first terms of $u_m$ are given in \cref{tab:u_n}.
\end{enumerate}

\section{Group action}
\label{sec:group}

We define the action by how its generators transform the basis elements of the space of correlation tensors:
\begin{equation}
  \begin{aligned}
    \be&\coloneqq e_{x_1}\otimes e_{x_2}\otimes e_{x_3}\otimes\ldots\otimes e_{x_N} \\
    g_1\cdot \be&\coloneqq e_{x_2}\otimes e_{x_1}\otimes e_{x_3}\otimes\ldots\otimes e_{x_N} \\
    g_2\cdot \be&\coloneqq e_{x_N}\otimes e_{x_1}\otimes e_{x_2}\otimes\ldots\otimes e_{x_{N-1}} \\
    g_3\cdot \be&\coloneqq e_{x_1+1}\otimes e_{x_2-1}\otimes e_{x_3}\otimes\ldots\otimes e_{x_N} \\
    g_4\cdot \be&\coloneqq e_{\bar{x}_1}\otimes e_{\bar{x}_2}\otimes e_{\bar{x}_3}\otimes\ldots\otimes e_{\bar{x}_N}
  \end{aligned}
  \label{eqn:generators}
\end{equation}
where $e_m\coloneqq -e_1$ and $e_{\bar{x}_n}\coloneqq -e_{m-x_n}$.
$g_1$ and $g_2$ generate the symmetric group permuting the parties~\cite{BBB+12}, $g_3$ encodes the periodic structure of the tensor, and $g_4$ its reflection structure.
We denote by $G$ the group generated by $g_1$, $g_2$, $g_3$, and $g_4$.
Note that we do not study the structure of $G$ itself, as we are only interested in its action on tensors defined in \cref{eqn:generators}.

The action defined by the generators preserves the deterministic strategies of \cref{eqn:deterministic}.
The orbits of this action on deterministic strategies will play an important role later.
We can accelerate their enumeration by exploiting the invariance under permutation of parties (given by $g_1$ and $g_2$) and the fact that $g_3$ cycles only two parties at a time.
This means that, for $N-1$ parties, we can restrict the enumeration of the strategies to the orbits of the subgroup generated by $g_3$, effectively using the last party to freely cycle the first $N-1$ parties.
This procedure gives a simple upper bound on the number of orbits
\begin{equation}
  2^mu_m^{N-1},
  \label{eqn:orbits_naive}
\end{equation}
which can be refined by ordering the orbits of the first $N-1$ parties to
\begin{equation}
  2^m\binom{u_m+N-2}{N-1},
  \label{eqn:orbits}
\end{equation}
where $u_m$ is the number of orbits of the subgroup generated by $g_3$, namely,
\begin{equation}
  u_m\coloneqq\frac{1}{2m}\sum_{2\nmid d\mid m}\varphi(d)2^{\frac{m}{d}}
  \label{eqn:u_n}
\end{equation}
where this formula comes from the enumeration of binary necklaces, see the On-Line Encyclopedia of Integer Sequences (\href{https://oeis.org/A000016}{A000016}) and references therein; see also \cref{tab:u_n} for the first terms of this sequence.
\cref{eqn:orbits} can be slightly improved, for instance, by exploiting $g_4$ (which we did in practice), but it is already good enough for our needs.

\begin{table*}[ht!]
  \centering
  \begin{tabular}{|c|c|c|c|c|c|c|c|c|c|c|c|c|c|c|c|c|c|c|c|c|c|}
    \hline
    $m$     & 2   & 3   & 4   & 5   & 6   & 7    & 8    & 9    & 10   & 11   & 12    & 13    & 14    & 15     & 16     & 17     & 18     & 19          & 20          & 21          \\ \hline
    ~$u_m$~ & ~1~ & ~2~ & ~2~ & ~4~ & ~6~ & ~10~ & ~16~ & ~30~ & ~52~ & ~94~ & ~172~ & ~316~ & ~586~ & ~1096~ & ~2048~ & ~3856~ & ~7286~ & ~$13\,798$~ & ~$26\,216$~ & ~$49\,940$~ \\ \hline
  \end{tabular}
  \caption{
    First values of the sequence $\{u_m\}_m$ defined in \cref{eqn:u_n}.
  }
  \label{tab:u_n}
\end{table*}

Next we study the Reynolds operator associated to the generic action on tensors, that is,
\begin{equation}
  \Gamma(\bp)\coloneqq\frac{1}{|G|}\sum_{g\in G}g\cdot\bp.
  \label{eqn:reynolds}
\end{equation}
By construction, this operator projects the correlation tensor $\bp$ onto a subspace of dimension $\lceil\tfrac{m}{2}\rceil$ such that the coordinates $\tilde{p}_{x_1\ldots x_N}$ of $\Gamma(\bp)$ satisfy
\begin{equation}
  \tilde{p}_{x_1\ldots x_N}=\left\{
    \begin{array}{ll}
      (-1)^{q_\mathrm{tot}}\tilde{p}_{r_\mathrm{tot},0\ldots0} & \text{if }r_\mathrm{tot}\leq\frac{m}{2} \\
      (-1)^{q_\mathrm{tot}+1}\tilde{p}_{m-r_\mathrm{tot},0\ldots0} & \text{if }r_\mathrm{tot}\geq\frac{m}{2}.
    \end{array}
  \right.
  \label{eqn:invariant_subspace}
\end{equation}
where $q_\mathrm{tot}$ and $r_\mathrm{tot}$ are the quotient and remainder of the Euclidean division of $\sum_nx_n$ by $m$.
Note that, when $m$ is even, the two conditions in \cref{eqn:invariant_subspace} are simultaneously fulfilled for $r_\mathrm{tot}=m/2$ so that the corresponding elements in the projected correlation tensor are 0.
This justifies the dimension $\lceil\tfrac{m}{2}\rceil$ given above.

Note that \cite[Appendix A]{Gis09} considers a similar type of Bell inequalities in the case where $N=2$ and the flip generator $g_4$ is omitted.
The construction of inequalities there was done by enumeration of all matrices respecting the symmetry, which was very soon intractable as $m$ grows.
In the following, we give two methods to generate multipartite Bell inequalities satisfying the symmetry: computing all facets of the symmetrised polytope (limited to small $m$) or deriving the one separating our point of interest given in \cref{eqn:GHZtensor} (via Frank-Wolfe algorithms).

\section{Symmetrised local polytopes}
\label{sec:polytope}

Following \cite{BGP10}, we first try to fully characterise the projected polytope, that is, $\Gamma(\mL)$.
In order to do so, we can enumerate all symmetrised deterministic strategies and then compute the facets of the resulting polytope.
The number of vertices found in the computationally tractable cases can be found in \cref{tab:vertices}.
For all these examples, the number of facets is equal to $2^{\lceil m/2\rceil}$.
This seems to indicate that the symmetric polytope is affinely equivalent to the cross-polytope in dimension $\lceil m/2\rceil$.
Although we considered the orbits giving rise to the extreme points of this polytope and tried to infer a generalisable pattern, we could not establish this fact.
We conjecture, however, that it holds in general, and hope that further research will identify these general extreme points.
Note that proving this may be motivated by the observation that symmetric facets seem to be facets of the non-symmetrised local polytope when $N$ is odd and $m$ even, so that obtaining them in general may directly give infinite families of facets for the local polytope.

\begin{table}[hb!]
  \centering
  \scalebox{0.9}{
    \begin{tabular}{|c|c|c|c|c|c|c|c|c|}
      \hline
      \backslashbox{$m\!\!\!\!\!\!$}{$\!\!\!\!\!\!N$}
         & 3            & 4            & 5           & 6      & 7           & 8      & 9     & 10    \\ \hline
      2  & 2            & 3            & 2           & 3      & 2           & 3      & 2     & 3     \\
      3  & 10           & 12           & 14          & 16     & 18          & 20     & 22    & 24    \\
      4  & 10           & 21           & 14          & 29     & 18          & 37     & 22    & 45    \\
      5  & 60           & 90           & 126         & 168    & 216         & 270    & ~330~ & ~396~ \\
      6  & 100          & 249          & 336         & 657    & 816         & ~1367~ &       &       \\
      7  & 640          & 1640         & 3740        & ~7774~ & ~$14\,990$~ &        &       &       \\
      8  & 1540         & 7889         & ~$22\,008$~ &        &             &        &       &       \\
      9  & $10\,032$    & $51\,260$    &             &        &             &        &       &       \\
      10 & $30\,494$    & ~$340\,349$~ &             &        &             &        &       &       \\
      11 & $243\,090$   &              &             &        &             &        &       &       \\
      12 & ~$799\,980$~ &              &             &        &             &        &       &       \\ \hline
    \end{tabular}
  }
  \caption{
    Number of vertices of the symmetrised local polytope, that is, number of different correlation tensors of the form \cref{eqn:invariant_subspace} obtained when applying the Reynolds operator \cref{eqn:reynolds} on all deterministic strategies defined in \cref{eqn:deterministic}.
  }
  \label{tab:vertices}
\end{table}

\section{Symmetric Frank-Wolfe}
\label{sec:fw}

As $m$ and $N$ grow, the enumeration of all symmetrised vertices soon becomes intractable, and \emph{a fortiori} the enumeration of all facets of the symmetrised polytope, see \cref{tab:vertices}.
Since our main interest lies in a specific facet, namely, the one separating the correlation tensor in \cref{eqn:GHZtensor}, we can use a different approach to get this exact facet.
The most natural reformulation leads to a linear programming (LP) instance, but it becomes intractable very soon~\cite{GGH+14,PBM+22}, essentially because enumerating all vertices of the (symmetrised) local polytope soon becomes impossible.
This problem has been addressed using the Gilbert algorithm~\cite{Gil66} adapted for quantum applications~\cite{BNV16}, and later reconnected to the corresponding field of constrained convex optimisation working on Frank-Wolfe algorithms~\cite{DIB+23}.
In this paper, we pursue this Frank-Wolfe approach, which allows us to move inside the local polytope without the need to enumerate its vertices, as only those which are relevant to our problem are explored.

More precisely, this class of first-order algorithms, also called conditional gradient methods~\cite{BCC+22}, uses a linear minimisation oracle (LMO) to find directions of progress within the feasible region and is quite memory efficient, which is suited for high-dimensional problems.
Moreover, in our case where the feasible region is a polytope, the decomposition of the final point returned by the algorithm is usually very sparse, which makes this approach a good way of obtaining the points defining the facet of interest.

Formally, we are solving
\begin{equation}
  \min\limits_{\bx\in\mL^{(m)}_N}\overbrace{\frac{1}{2}\big\|\bx-v_0\br^{(m)}_N\big\|_2^2}^{f(\bx)},
  \label{eqn:fw}
\end{equation}
where the choice of the initial visibility $v_0$ will be discussed below (take $v_0=1$ to start with).
The core idea of the original Frank-Wolfe algorithm~\cite{FW56,Jag13} is to repeatedly move towards the minimiser of the so-called LMO, which is the linearisation of our function at the current iterate $\bx_t$:
\begin{equation}
  \bv_t\coloneqq\arg\min\limits_{\bx\in\mL^{(m)}_N}\big\langle\overbrace{\bx_t-v_0\br^{(m)}_N}^{\nabla f(\bx_t)},\bx\big\rangle.
  \label{eqn:lmo}
\end{equation}
In our case, this oracle amounts to finding a deterministic strategy reaching the local bound of the Bell inequality defined by the gradient at the current iterate.
We explain in more detail how this is done in \cref{sec:bound}.
Since this LMO can be costly, variants of this algorithm have been developed that feature a memory (active set) used to recycle previously found vertices and hence reduce the number of calls to the LMO.
Moreover, for high $m$, we employ a heuristic LMO.
We refer to~\cite{DIB+23} and references therein for more details on the method and only comment here on the choice of $v_0$.

As already mentioned, we start with $v_0=1$, which gives a first separating hyperplane, and then use the corresponding visibility (that is, $v$ such that $v\br^{(m)}_N$ lies on this hyperplane) to start the algorithm again, keeping the active set, i.e., the deterministic strategies already found.
Since we are interested in facets of the symmetrised polytope, we repeat this procedure until the number of points in the active set reaches the dimension of the search space, which is conveniently quite small here.

In our symmetric case, we can indeed navigate in the reduced space of dimension $\lceil\tfrac{m}{2}\rceil$, keeping in mind that the scalar product has to be properly weighted in order to reflect the actual geometry of the initial tensors.
What allows us to perform this dimension reduction is the fact, mentioned above in \cref{sec:polytope}, that symmetrised vertices can be viewed both geometrically --- as the projection on the symmetric subspace --- and algebraically --- as the sum of the orbit of the atom found by the LMO, see \cref{eqn:reynolds}.
Phrased differently, this means that every atom found is virtually added to the active set together with its entire orbit, placing the same weight on all these symmetrically equivalent atoms.

Although the dimension of the space is independent of $N$, the complexity of the LMO depends on $N$ so that it becomes more and more expensive as this number increases.
More precisely, the heuristic approach used in~\cite{DIB+23} (alternating minimisation) acts on full tensors (of size $m^N$) and requires more and more iterations to converge when $N$ grows.
One can naturally avoid storing these full tensors in memory, but the complexity still scales quite poorly.
We nonetheless underline that this LMO is called very infrequently in the version of Frank-Wolfe that we use here (with active set)~\cite{BPTW18}.
Typically the total number of calls is a few times the dimension $\lceil\tfrac{m}{2}\rceil$, hence very reasonable; for example, in the instance with $N=3$ and $m=224$ given in \cref{tab:GHZsummary}, it was called 539 times.

Moreover, as $N$ grows, the enumeration of orbits discussed below becomes more and more competitive with respect to the alternating heuristic just mentioned.
For $N\geq8$ we systematically use this enumeration as it is faster on top of being exact.

\section{Local bounds}
\label{sec:bound}

The question of computing local bounds is notoriously complex.
Already for $N=2$, it is a quadratic unconstrained binary optimisation (QUBO) instance, and the degree of the problem increases with $N$.
One could reformulate these problems into linear ones at the expense of increasing the number of variables~\cite{CELR22}, but such a reformulation is out of the question in practice as the resulting problems are far too big, already for $N=2$.
In this section, we explain how symmetry can significantly reduce the naive enumeration of cases.

Formally, for an $N$-partite tensor $\bbf$ defining a Bell inequality, the problem reads
\begin{equation}
  \max_{\va^{(1)}\ldots\va^{(N)}}\sum_{\vx\in[m]^N}\!\!f_{\vx} \prod_{n=1}^Na^{(n)}_{x_n}
  \label{eqn:L}
\end{equation}
where the maximisation is performed for $\va^{(n)}\in\{\pm1\}^m$, that is, over deterministic strategies.

In general, without symmetries, it suffices to enumerate $2^{(N-1)(m-1)}$ strategies.
This is because the optimal strategy for the last party is fixed when the other $N-1$ parties have chosen theirs, as follows:
\begin{equation}
  a^{(N)}_{x_N}=\mathrm{sign}\left(\sum_{\vx\in[m]^{N-1}}\!\!\!\!\!f_{\vx} \prod_{n=1}^{N-1}a^{(n)}_{x_n}\right),
  \label{eqn:sign}
\end{equation}
where we set $\mathrm{sign}(0)=1$.
Moreover, since we do not consider marginals here, we can fix $a_1^{(n)}=-1$ for $n\in[N-1]$, up to flipping the signs of both $\va^{(n)}$ and $\va^{(N)}$.

With symmetries, we can further restrict to the orbits of the deterministic strategies.
Since the last strategy is fixed as in the nonsymmetric case, see \cref{eqn:sign}, the factor $2^m$ in \cref{eqn:orbits} can be dropped to obtain an upper bound on the number of cases to be considered:
\begin{equation}
  \binom{u_m+N-2}{N-1},
  \label{eqn:enumeration}
\end{equation}
where we recall that $u_m$ is defined in \cref{eqn:u_n} and its first terms are given in \cref{tab:u_n}.
In practice, we used GAP~\cite{GAP} to obtain the orbits of interest (up to $m=26$).

\section{Results}
\label{sec:results}

\begin{figure}[ht!]
  \centering
  \vspace{30pt}%
  \hspace{-4pt}%
  \begin{tikzpicture}
    \begin{axis}[
        xmin=4, xmax=26,
        ymin=0.49, ymax=0.508,
        xtick={2, 4, 8, 12, 16, 20, 24},
        ytick={0.49, 0.5, ..., 0.508},
        yticklabel style={/pgf/number format/.cd, fixed, precision=3},
        legend pos=north east,
        ymajorgrids=true,
        grid style=dashed,
        width=4.6cm,
        height=4.5cm,
      ]
      \addplot[color=darkblue,mark=o,mark size=1.5pt]
      coordinates{
        (2,  0.5                )
        (3,  0.5714285714285714 )
        (4,  0.5                )
        (5,  0.5079365079365079 )
        (6,  0.49505300219518955)
        (7,  0.4991077277438634 )
        (8,  0.4931688508372498 )
        (9,  0.49675865746496567)
        (10, 0.49206349206349204)
        (11, 0.49288015789803835)
        (12, 0.49220761333287416)
        (13, 0.49248876265628627)
        (14, 0.49175803308261706)
        (15, 0.4931950144310966 )
        (16, 0.49159914189635234)
        (17, 0.491920012598196  )
        (18, 0.4917984589957535 )
        (19, 0.491835751627009  )
        (20, 0.49152650754250965)
        (21, 0.4920977794255365 )
        (22, 0.4914274507842722 )
        (23, 0.4915536046536135 )
        (24, 0.49149952411488934)
        (25, 0.49161596587120765)
        (26, 0.49143412002270725)
      };
      \legend{$N=3$}
    \end{axis}
  \end{tikzpicture}%
  \hspace{14pt}%
  \begin{tikzpicture}
    \begin{axis}[
        xmin=4, xmax=21,
        ymin=0.32, ymax=0.355,
        xtick={2, 4, 8, 12, 16, 20},
        ytick={0.32, 0.33, ..., 0.355},
        yticklabel style={/pgf/number format/.cd, fixed, precision=3},
        legend pos=north east,
        ymajorgrids=true,
        grid style=dashed,
        width=4.6cm,
        height=4.5cm,
      ]
      \addplot[color=darkblue,mark=o,mark size=1.5pt]
      coordinates{
        (2,  0.5                )
        (3,  0.3902439024390244 )
        (4,  0.3535533905932738 )
        (5,  0.34293421901788673)
        (6,  0.33800341115120336)
        (7,  0.33302019360999857)
        (8,  0.33059451439907184)
        (9,  0.3299447531561415 )
        (10, 0.3283612799754399 )
        (11, 0.3274033734808601 )
        (12, 0.3272115933660127 )
        (13, 0.32636857140563713)
        (14, 0.3260611921443605 )
        (15, 0.32607481350481454)
        (16, 0.3254843462017326 )
        (17, 0.3252998973283111 )
        (18, 0.3253094988463458 )
        (19, 0.3250044561285162 )
        (20, 0.3249317496789528 )
        (21, 0.32493632605237044)
      };
      \legend{$N=4$}
    \end{axis}
  \end{tikzpicture}\\[20pt]
  \hspace{2pt}%
  \begin{tikzpicture}
    \begin{axis}[
        xmin=4, xmax=18,
        ymin=0.205, ymax=0.226,
        xtick={2, 4, 8, 12, 16},
        ytick={0.21, 0.22, ..., 0.226},
        yticklabel style={/pgf/number format/.cd, fixed, precision=3},
        legend pos=north east,
        ymajorgrids=true,
        grid style=dashed,
        width=4.6cm,
        height=4.5cm,
      ]
      \addplot[color=darkblue,mark=o,mark size=1.5pt]
      coordinates{
        (2,  0.25               )
        (3,  0.26229508196721313)
        (4,  0.21875            )
        (5,  0.2252079334612924 )
        (6,  0.21295771228877922)
        (7,  0.21665772357746152)
        (8,  0.2107686819907335 )
        (9,  0.21346735774428915)
        (10, 0.20985284708893154)
        (11, 0.2115819734792637 )
        (12, 0.20937680653908505)
        (13, 0.21061766035352042)
        (14, 0.20903608705234783)
        (15, 0.21009874174071508)
        (16, 0.20883755860372183)
        (17, 0.20961165980071392)
        (18, 0.20872147429211252)
      };
      \legend{$N=5$}
    \end{axis}
  \end{tikzpicture}%
  \hspace{14pt}%
  \begin{tikzpicture}
    \begin{axis}[
        xmin=4, xmax=16,
        ymin=0.13, ymax=0.155,
        xtick={2, 4, 8, 12, 16},
        ytick={0.13, 0.14, ..., 0.155},
        yticklabel style={/pgf/number format/.cd, fixed, precision=3},
        legend pos=north east,
        ymajorgrids=true,
        grid style=dashed,
        width=4.6cm,
        height=4.5cm,
      ]
      \addplot[color=darkblue,mark=o,mark size=1.5pt]
      coordinates{
        (2,  0.25               )
        (3,  0.17534246575342466)
        (4,  0.15467960838455727)
        (5,  0.14653363089066254)
        (6,  0.14236048464290246)
        (7,  0.13970738913563152)
        (8,  0.13809069699042895)
        (9,  0.13705100676301366)
        (10, 0.1362269294168349 )
        (11, 0.13564751197141783)
        (12, 0.1352409235208644 )
        (13, 0.13487600958564608)
        (14, 0.13460989394551465)
        (15, 0.13441384138034662)
        (16, 0.13421819063819734)
      };
      \legend{$N=6$}
    \end{axis}
  \end{tikzpicture}\\[20pt]
  \begin{tikzpicture}
    \begin{axis}[
        xmin=4, xmax=12,
        ymin=0.085, ymax=0.096,
        xtick={2, 4, 6, 8, 10, 12},
        ytick={0.085, 0.09, ..., 0.096},
        legend pos=north east,
        ymajorgrids=true,
        grid style=dashed,
        width=4.6cm,
        height=4.5cm,
      ]
      \addplot[color=darkblue,mark=o,mark size=1.5pt]
      coordinates{
        (2,  0.125              )
        (3,  0.1170018281535649 )
        (4,  0.09375            )
        (5,  0.09502598366740905)
        (6,  0.08866073818981952)
        (7,  0.08982615672158163)
        (8,  0.08691021037531904)
        (9,  0.08778609785058828)
        (10, 0.08611727613289011)
        (11, 0.0867515476772798 )
        (12, 0.08569184124708779)
      };
      \legend{$N=7$}
    \end{axis}
  \end{tikzpicture}%
  \hspace{12pt}%
  \begin{tikzpicture}
    \begin{axis}[
        xmin=4, xmax=11,
        ymin=0.055, ymax=0.067,
        xtick={2, 4, 6, 8, 10},
        ytick={0.055, 0.06, ..., 0.067},
        legend pos=north east,
        ymajorgrids=true,
        grid style=dashed,
        width=4.6cm,
        height=4.5cm,
      ]
      \addplot[color=darkblue,mark=o,mark size=1.5pt]
      coordinates{
        (2,  0.125               )
        (3,  0.07802499238037183 )
        (4,  0.06629126073623884 )
        (5,  0.061547881988980886)
        (6,  0.05912517281887531 )
        (7,  0.05769767593179221 )
        (8,  0.05679614646630232 )
        (9,  0.0561899019976151  )
        (10, 0.05575483095683277 )
        (11, 0.05543730560853126 )
      };
      \legend{$N=8$}
    \end{axis}
  \end{tikzpicture}\\[20pt]
  \begin{tikzpicture}
    \begin{axis}[
        xmin=4, xmax=10,
        ymin=0.034, ymax=0.041,
        xtick={2, 4, 6, 8, 10},
        ytick={0.034, 0.036, ..., 0.041},
        legend pos=north east,
        ymajorgrids=true,
        grid style=dashed,
        width=4.6cm,
        height=4.5cm,
      ]
      \addplot[color=darkblue,mark=o,mark size=1.5pt]
      coordinates{
        (2,  0.0625              )
        (3,  0.052021946758788865)
        (4,  0.0400390625        )
        (5,  0.039845867598915796)
        (6,  0.036781001712260175)
        (7,  0.037048231303264754)
        (8,  0.03569784985378794 )
        (9,  0.035957756776986255)
        (10, 0.03520714954320465 )
      };
      \legend{$N=9$}
    \end{axis}
  \end{tikzpicture}%
  \hspace{12pt}%
  \begin{tikzpicture}
    \begin{axis}[
        xmin=4, xmax=9,
        ymin=0.022, ymax=0.029,
        xtick={2, 4, 6, 8, 10},
        ytick={0.022, 0.024, ..., 0.029},
        legend pos=north east,
        ymajorgrids=true,
        grid style=dashed,
        width=4.6cm,
        height=4.5cm,
      ]
      \addplot[color=darkblue,mark=o,mark size=1.5pt]
      coordinates{
        (2, 0.0625              )
        (3, 0.034682472480948355)
        (4, 0.028311892606102004)
        (5, 0.02579155673061677 )
        (6, 0.024521498310264917)
        (7, 0.023786247084182498)
        (8, 0.023321701257653968)
        (9, 0.023008858361526673)
      };
      \legend{$N=10$}
    \end{axis}
  \end{tikzpicture}
  \caption{
    Critical visibility $v_m^{\GHZ_N}$ for the nonlocality of the $N$-partite GHZ state with each party performing $m$ measurements forming a regular polygon on the XY plane.
    These visibilities naturally give upper bounds on the nonlocality robustness of the $N$-partite GHZ state, see \cref{eqn:vc,eqn:vcup}.
    We refer to the supplemental file for the analytical values and corresponding Bell inequalities and to \cref{tab:GHZup} for approximate values.
    The non-monotonic behaviour is related to the non-optimality of the quantum strategy chosen to witness the nonlocality of the $N$-partite GHZ state for $m$ measurements per party.
    We expect optimal strategies to enjoy similar symmetries to the ones exploited here but do not study this point in this work.
  }
  \label{fig:GHZ}
\end{figure}
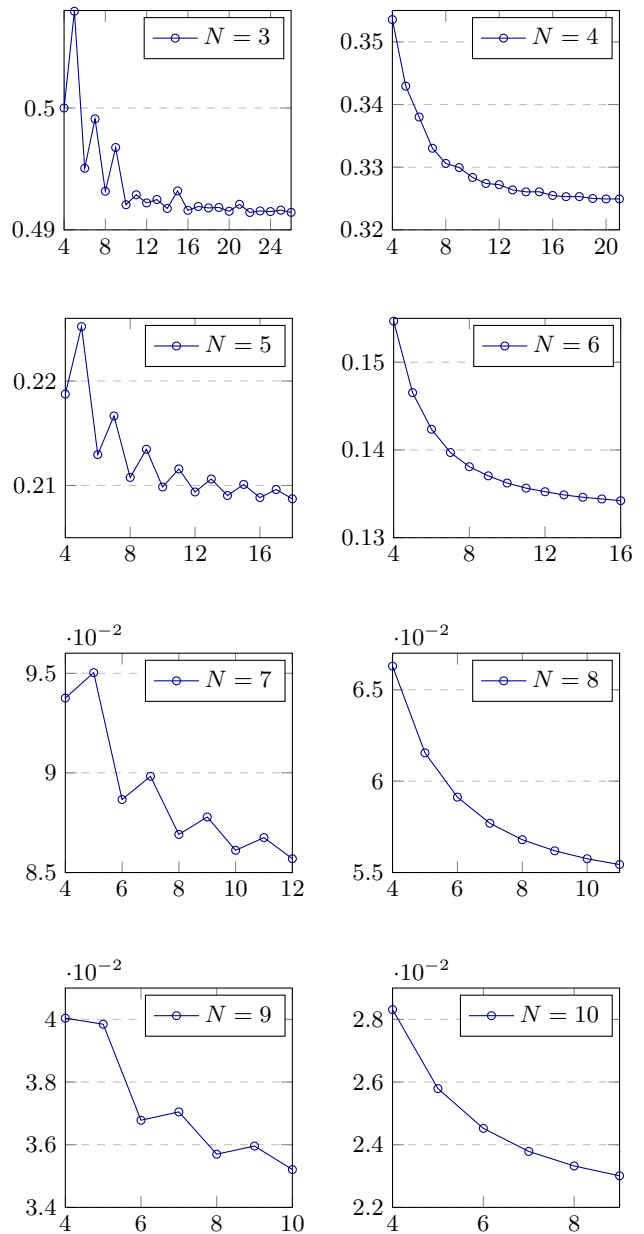

We are now ready to present the robustnesses obtained by using the method presented in the previous sections.
The evolution of the value of $v_m^{\GHZ_N}$ is shown in \cref{fig:GHZ} and we summarise the best values in \cref{tab:GHZsummary} together with some values relevant for the discussion in \cref{sec:consequences}.
All Bell inequalities and closed-form expressions of the values given here can be found in the supplemental text file accompanying this article.

Let us show an elegant example to give a flavour of the kind of inequalities found by our method.
We pick $N=5$ and $m=10$, so that the facet $\bbf$ derived with our procedure and given in the supplemental file has its first $m$ elements $f_{x_10000}$ for $x_1\in[m]$ being
\begin{equation}\nonumber
  [988,\ 0,\ 575,\ 0,\ -575,\ 0,\ 575,\ 0,\ -575,\ 0],
\end{equation}
all the others being deduced by symmetry, see \cref{eqn:invariant_subspace}.
With a slight improvement on \cref{eqn:enumeration} obtained by using $g_4$ from \cref{eqn:generators}, we obtain the upper bound $242\,873$ on the number of orbits.
By enumerating only this many cases instead of all $2^{36}\approx7\times10^{10}$ strategies, we get the local bound $L_5^{(10)}=3\,280\,000$.
The quantum value reached by the GHZ state is $Q_5^{(10)}=15\,630\,000$, which finally gives $v_{10}^{\GHZ_5}=L_5^{(10)}/Q_5^{(10)}\approx0.20985$.

One immediate comment when considering \cref{fig:GHZ} is the non-monotonicity of the curves when $N$ is odd.
This happens because the measurements used are the same on all parties, which is not optimal in terms of robustness.
For instance, for $N=3$, we could use the following asymmetric setting: the first and third parties perform the same measurements given by a regular polygon in the XY plane, and the second uses the same polygon, but rotated by an angle $\pi/(2m)$.
The resulting correlation tensor also enjoys symmetries, but they are slightly different, although the very same method as the one presented in this article can be equally successfully applied.

Interestingly, the integer coefficients of the inequalities found become quite large when $m$ increases (for instance, 14 digits for $N=3$ and $m=24$).
Although this property is not surprising from an algebraic point of view (they are hyperplanes going through integer vertices in an increasing dimension), most inequalities studied in the literature contain small integers, and the author of~\cite{Gis09} also restricted the search of symmetric inequalities to such small values.
Maybe a close examination of the inequalities constructed in this work (or similar ones that the authors can compute on demand) could reveal an interesting method to build general inequalities of interest.

As far as the computation time is concerned, we ran all instances on a 64-core Intel$^\circledR$ Xeon$^\circledR$ Gold 6338 machine and the longest one was for $N=10$ and $m=9$, which took ten days.
For $N\leq6$, however, the running time is more reasonable: from a few seconds for $m\leq10$ to a few hours beyond.
Typically, the case $N=3$ and $m=224$ took 3 hours.

\begin{table*}
  \centering
  \begin{tabular}{|c|c|c|c|c|c|c|c|c|c|}
    \hline
    $N$                                           & 3         & 4         & 5         & 6         & 7         & 8         & 9         & 10        & Comments                   \\ \hline
    ~$2^{(1-N)/2}$~                               & 0.5       & 0.35355   & 0.25      & 0.17678   & 0.1250    & 0.08839   & 0.0625    & 0.04419   & Mermin \cite{Mer90}        \\ \hline
    $m=4$                                         & 0.5       & 0.35355   & 0.21875   & 0.15468   & 0.09375   & 0.06629   & 0.04004   & 0.02831   & \cref{sec:m=4}             \\ \hline
    $m$                                           & 22        & 20        & 18        & 16        & 12        & 11        & 10        & 9         &                            \\
    $v_m^{\GHZ_N}$                                & ~0.49143~ & ~0.32493~ & ~0.20872~ & ~0.13422~ & ~0.08569~ & ~0.05544~ & ~0.03521~ & ~0.02301~ &                            \\ \hline
    $m$                                           & 224       & 128       & 64        & 32        & 16        &           &           &           &                            \\
    $v_m^{\GHZ_N}$                                & 0.49132   & 0.32384   & 0.20824   & 0.13326   & 0.08526   &           &           &           & \multirow{-2}{*}{Putative} \\ \hline
    $v_{XY}^{\GHZ_N}\geq$                         & 0.49129   & 0.32374   & 0.20793   & 0.13231   & 0.08243   & 0.05108   & 0.03149   & 0.01974   & ~\cref{sec:vXY}~           \\ \hline
    $v_{\mathrm{low}}^N$                          & 0.32489   & 0.22335   & 0.15354   & 0.10555   & 0.07256   & 0.04989   & 0.03429   & 0.02358   & \cref{sec:activation}      \\ \hline
  \end{tabular}
  \caption{
    Critical visibility $v_m^{\GHZ_N}$ for the nonlocality of the $N$-partite GHZ state with each party performing $m$ measurements forming a regular polygon on the XY plane.
    We also include robustnesses obtained with a large number of measurements for which computing the local bound is out of reach, even with our symmetric approach, see \cref{tab:GHZup} for more values.
    The lower bound on $v_{XY}^{\GHZ_N}$ (see \cref{sec:vXY}) is, however, rigorously proven as it only requires a valid local model.
  }
  \label{tab:GHZsummary}
\end{table*}

\newpage

\section{Consequences}
\label{sec:consequences}

\subsection{Extension of the inequality for \texorpdfstring{$m=4$}{m=4} to all \texorpdfstring{$N$}{N}}
\label{sec:m=4}

For $m=4$, the inequalities found with our method have a very simple form: $[1, 0, 0, 0]$ for odd $N$ and $[0, 1, 0, -1]$ for even $N$, straightforwardly reaching the quantum value $Q_N^{(4)}$ of $4^{N-1}$ and $4^{N-1}\sqrt2$, respectively.
However, the computation of the local bound $L_N^{(4)}$, whose first values are given in \cref{tab:m=4}, is not as simple.
In \cref{app:m=4}, we show that $L_{N+4}^{(4)}=8\left(L_{N+2}^{(4)}-L_N^{(4)}\right)$ and $L_{2n}^{(4)}=4L_{2n-1}^{(4)}$ so that the subsequences of this local bound form Lucas sequences.
From there it follows that:
\begin{equation}
  L_{2n-1}^{(4)}=\frac{4^{n-1}l_n}{\sqrt2}\quad\text{and}\quad L_{2n}^{(4)}=\frac{4^nl_n}{\sqrt2}
  \label{eqn:m=4}
\end{equation}
where
\begin{equation}
  l_n\coloneqq\left(1+\frac{1}{\sqrt2}\right)^n-\left(1-\frac{1}{\sqrt2}\right)^n.
  \label{eqn:ln}
\end{equation}

The gist of the proof of the validity of this local bound is the following: there are only $u_4=2$ orbits per party, so that the total number of strategies to check is $N$, see \cref{eqn:enumeration}.
In this example, the best two strategies reaching the local bound are the ones in which all $N-1$ first parties use the same deterministic strategy (out of the two orbits available).
Note that, in this case, as the dimension of the space is also $\lceil\tfrac{m}{2}\rceil=2$, we can extract the explicit local model of $v_4^{\GHZ_N}\br_N^{(4)}$ as in \cref{eqn:fw_example}.

Similar inequalities were already mentioned in~\cite{BBT05} but the computation of the local bound, which did not use any symmetry reduction, was only done up to $N=5$.
Although the proof in \cref{app:m=4} is admittedly tedious, we expect further research to simplify the argumentation and to generalise it to more measurements.
A motivation to investigate in this direction is the observation that the trivial strategy is saturating all inequalities given in this work, see \cref{sec:vXY} below.

\begin{table*}[ht!]
  \centering
  \begin{tabular}{|c|c|c|c|c|c|c|c|c|c|c|c|c|c|c|c|}
    \hline
    $N$           & 3   & 4    & 5    & 6     & 7     & 8      & 9      & 10          & 11          & 12          & 13           & 14           & 15           & 16              & 17              \\ \hline
    ~$L_N^{(4)}$~ & ~8~ & ~32~ & ~56~ & ~224~ & ~384~ & ~1536~ & ~2624~ & ~$10\,496$~ & ~$17\,920$~ & ~$71\,680$~ & ~$122\,368$~ & ~$489\,472$~ & ~$835\,584$~ & ~$3\,342\,336$~ & ~$5\,705\,728$~ \\ \hline
  \end{tabular}
  \caption{
    Local bounds of the inequalities defined by $[1, 0, 0, 0]$ for odd $N$ and $[0, 1, 0, -1]$ for even $N$, see \cref{eqn:m=4}.
  }
  \label{tab:m=4}
\end{table*}

\subsection{Lower bounds on the nonlocality robustness with all projective measurements in the XY plane}
\label{sec:vXY}

In the qubit case, the visibility of $1/\sqrt2$ does not only correspond to the noise threshold reached by CHSH, but also to $1/K_G(2)$, that is, the best robustness achievable with measurements lying on a great circle of the Bloch sphere~\cite{AGT06}.
With our method we can obtain bounds on generalisations of this number: $v_m^{\GHZ_N}$ can indeed be turned into a lower bound on the nonlocality threshold $v_{XY}^{\GHZ_N}$ of the $N$-partite GHZ state under projective measurements in the XY plane.
More precisely,
\begin{equation}
  v_{XY}^{\GHZ_N}\geq\left[\cos\left(\frac{\pi}{2m}\right)\right]^Nv_m^{\GHZ_N},
  \label{eqn:vXY}
\end{equation}
where the trigonometric factor is the shrinking factor of the regular polygon with $2m$ vertices.

Note that valid lower bounds on $v_m^{\GHZ_N}$ can safely be plugged in \cref{eqn:vXY}, and that our algorithm can easily produce such lower bounds.
The main bottleneck to obtain $v_m^{\GHZ_N}$ is the computation of the local bound, which soon becomes intractable, although we have considerably pruned the number of strategies to consider.
But, in order to obtain a lower bound $v$ on $v_m^{\GHZ_N}$, we only need an explicit local model of $v\br_N^{(m)}$, that is, a convex decomposition of this point in terms of deterministic strategies; this is precisely what the Frank-Wolfe algorithm produces.
Moreover, given the small number of symmetrised vertices involved in the final decomposition, we can retrieve the \emph{putative} facet of the symmetrised local polytope.
This wording underlines the fact that we are unable to rigorously establish the corresponding local bound and hence the property of indeed being a facet (and not simply a separating hyperplane containing $\lceil\tfrac{m}{2}\rceil$ deterministic strategies).
Interestingly, as mentioned above, the strategy consisting of always answering $-1$ on all $N-1$ first parties is always reaching the value given by our heuristic, which seems to indicate that the geometry of the problem could be leveraged to prove the validity of the heuristic local bound.
We could not use it to our advantage and leave this question open for further work.

With the local model obtained above for $m=4$, we can also derive bounds on $v_{XY}^{\GHZ_N}$ for all $N$ by combining \cref{eqn:m=4,eqn:vXY}.
Contrary to those derived with Mermin's inequality, these bounds are higher than the entanglement threshold~\cite{DC00}.

\subsection{Detection efficiency}


We consider the case where the parties share an $N$-partite GHZ state in \cref{eqn:GHZ} and detect the particles with the same efficiency $\eta$.
When the particle is not detected, the parties agree to output $+1$; this ensures that the measurements remain dichotomic.
Note, however, that the parties could also use a third outcome in the case of no detection~\cite{MP03,CC19}.
If all detectors fire, which happens with a probability of $\eta^N$, then the $N$-partite full correlation Bell inequality $I$ can be maximally violated, that is, we have $I=Q_N^{(m)}$, the quantum value.
If only some of the detectors fire, we have $I=0$, since the inequality $I$ contains no marginal terms.
On the other hand, if no detector fires, which happens with a probability of $(1-\eta)^N$, the local bound can be reached, i.e., we have $I=L_N^{(m)}$.
Consequently, the entire data violate the inequality $I$ if and only if
\begin{equation}
  \eta^N Q_N^{(m)} + (1-\eta)^N L_N^{(m)} > L_N^{(m)},
  \label{eqn:criticaleff}
\end{equation}
and then dividing by $L_N^{(m)}$ to make $v_m^{GHZ_N}=L_N^{(m)}/Q_N^{(m)}$ appear, we arrive at
\begin{equation}
  \frac{\eta_\mathrm{crit}^N}{v_m^{\GHZ_N}}+(1-\eta_\mathrm{crit})^N=1,
  \label{eqn:efficiency}
\end{equation}
which defines the critical detection efficiency threshold $\eta_\mathrm{crit}$ for $N$ parties and $m$ measurements.

In~\cite{KLV18}, the value of $\eta_\mathrm{crit}=0.7706$ was calculated (based on the WWWK\.ZB inequalities, see references therein) for the special case of $N=4$ and $m=2$.
This value is already reproduced by our inequality for $N=4$ and $m=4$, and we steadily improve on this critical efficiency when $m$ increases, reaching, for $N=4$ and $m=19$, the value $\eta_\mathrm{crit}=0.7544$.
More generally, the bank of inequalities provided in this work should be of interest as they all feature good detection efficiency with few measurements.
Naturally, these critical efficiencies get better as the number of parties increases~\cite{BHMR03}.

\subsection{Activation of nonlocality in star networks}
\label{sec:activation}


Consider a star network in which a central party shares two-qubit isotropic states with visibility $v$ with $N$ surrounding parties.
Upon projection onto the GHZ state by this central party, the state shared by the surrounding parties reads
\begin{equation}
  \rho_N = v^N\ketbra{\GHZ_N}+(1-v^{N-1})\frac{\id_{2^N}}{2^N}+\rho_{\mathrm{rest}},
  \label{eqn:rest}
\end{equation}
where $\rho_{\mathrm{rest}}$ vanishes when each party performs measurements on the XY plane, so that we are effectively left with the correlations of an $N$-partite GHZ state with noise $v^N$.

In~\cite{CASA11} this entanglement swapping procedure was used to show that nonlocality can be activated for $N\geq21$.
This means that there is a visibility $v$ for which the initial two-qubit states are local, but where nonlocality can be demonstrated in the star network using the method described above.
As a first remark, the proof used in~\cite{CASA11} relied on the lower bound $0.6595$ on the nonlocality threshold of the two-qubit isotropic state.
As this bound has been improved since then to $v_{\mathrm{low}}\approx0.6875$~\cite{DIB+23}, the number of parties for which nonlocality activation occurs is $N=10$.
Indeed, for this number of parties we have
\begin{equation}
  \left(\frac{2}{\pi}\right)2^{\frac1N} < v_{\mathrm{low}},
\end{equation}
which shows that the inequality from \cite{SSB+05}, with an infinite number of measurements, can be used to witness the activation of nonlocality in networks.

Our results strengthen this result though.
For $N=10$ and $m=8$ or $m=9$, we indeed have
\begin{equation}
  \left(v_m^{\GHZ_N}\right)^{\frac1N} < v_{\mathrm{low}},
\end{equation}
which shows that the activation of nonlocality in networks can be demonstrated with a finite (and fairly small) number of measurements.

\section{Conclusion}

In this article we studied the nonlocality robustness of the $N$-partite GHZ state through a specific quantum strategy where all parties perform measurements forming a regular polygon in the XY plane of the Bloch sphere.
The specificity of this setting is motivated by the symmetry of the resulting correlation tensor and the quality of the corresponding critical visibilities, giving bounds on the nonlocality robustness.
Studying these symmetries allows us to devise efficient ways of finding tight Bell inequalities for which the local bound can also be computed thanks to symmetries.
The largest instance we solve is $N=10$ and $m=9$, which would feature, without symmetrisation, correlation tensors with $3.5\times10^9$ elements and would require enumerating $4.7\times10^{21}$ deterministic strategies; instead, we only need $5$ elements to represent the correlation tensors and iterate over $1.4\times10^8$ orbits to obtain the local bound.
These critical visibilities have some immediate consequences in terms of detection efficiency and activation of nonlocality in star networks, but they should essentially be seen as a proof of concept that symmetries can be leveraged in the main algorithmic ingredient of this work: Frank-Wolfe algorithms.

Finding other relevant symmetric cases would be a natural next step; in particular, identifying a bipartite correlation tensor featuring nice symmetries and a good robustness would be an excellent way to improve on the lower bound on the Grothendieck constant of order three.
But more generally, showcasing the use of symmetrisation in this context may be taken as an inspiration for all contexts in which Frank-Wolfe algorithms may play a role, e.g., entanglement detection, inflation in networks, and large-scale semidefinite programming.

\section{Acknowledgements}

The authors are grateful to Mathieu Besançon, Patrick Gelß, Gabriele Iommazzo, Sebastian Knebel, and Marc-Olivier Renou for various discussions.
This research was partially funded by the DFG Cluster of Excellence MATH+ (EXC-2046/1, Project No.~390685689) funded by the Deutsche Forschungsgemeinschaft (DFG).
T.~V.~acknowledges the support of the European Union (QuantERA eDICT) and the National Research, Development and Innovation Office NKFIH (Grants No.~2019-2.1.7-ERA-NET-2020-00003 and No.~K145927).
GPT 6 Astra was used to correct minor grammatical and mathematical errors.

\bibliography{DVP24}

\onecolumngrid
\appendix

\section{Computation of the local bound for \texorpdfstring{$m=4$}{m=4}}
\label{app:m=4}

Here we outline the proof of the local bound in the case of an odd number of parties, that is, for the facet $[1, 0, 0, 0]$.
The other case, namely, $[0, 1, 0, -1]$ for an even number of parties, can be treated similarly.

We start by defining the matrices
\begin{equation}\label{eqn:RS}
  R\coloneqq
  \begin{pmatrix}
     1 &  1 &  1 & 1 \\
    -1 &  1 &  1 & 1 \\
    -1 & -1 &  1 & 1 \\
    -1 & -1 & -1 & 1
  \end{pmatrix}
  =V
  \begin{pmatrix}
    \alpha_{++} & & & \\
    & \!\!\!\!\!\alpha_{-+} & & \\
    & & \!\!\!\!\!\alpha_{+-} & \\
    & & & \!\!\!\!\!\alpha_{--}
  \end{pmatrix}
  V^\dagger
  \quad\text{and}\quad
  S\coloneqq
  \begin{pmatrix}
     1 &  1 & -1 &  1 \\
    -1 &  1 &  1 & -1 \\
     1 & -1 &  1 &  1 \\
    -1 &  1 & -1 &  1
  \end{pmatrix}
  =V
  \begin{pmatrix}
    \alpha_{+-} & & & \\
    & \!\!\!\!\!\alpha_{--} & & \\
    & & \!\!\!\!\!\alpha_{++} & \\
    & & & \!\!\!\!\!\alpha_{-+}
  \end{pmatrix}
  V^\dagger,
\end{equation}
which we directly diagonalised with the following eigenvalues and eigenvectors:
\begin{equation}
  \alpha_{\pm\pm}=1\pm\left(\sqrt2\pm1\right)\mi
  \quad\text{and}\quad
  V=\frac12
  \begin{pmatrix}
    \omega^{-3} & \omega^3 & \omega^{-1} & \omega \\
    -\mi        & \mi      & \mi         & -\mi \\
    \omega^{-1} & \omega   & \omega^{-3} & \omega^3 \\
    1           & 1        & 1           & 1
  \end{pmatrix}
  ,\quad\text{where}\quad
  \omega=\me^{\mi\pi/4}.
\end{equation}
Note that $V$ is, up to trivial operations, a Vandermonde matrix.

The reason to introduce these matrices is that
\begin{equation}\label{eqn:Lij}
  L_{i,j}\coloneqq\|R^iS^j\|_1
\end{equation}
is exactly the value attained in a scenario with $i+j+1$ parties, when $i$ of them choose the trivial strategy $(1, 1, 1, 1)$ and $j$ of them the strategy $(1, 1, -1, 1)$.
Since there are $m=4$ measurements, these two possibilities are the only two orbits.
Note that the order of the parties does not matter thanks to the symmetries, which can be seen in the commutation of $R$ and $S$.
Actually, only the first column of $R^iS^j$ truly matters in the definition of $L_{i,j}$ since the Bell inequality considered here is $[1, 0, 0, 0]$, but, normalising the 1-norm (so that it is submultiplicative) and noting that all columns have the same 1-norm by symmetry, \cref{eqn:Lij} is equivalent to this.

Having diagonalised the matrices $R$ and $S$ allows us to explicitly derive the expression of $L_{i,j}$:
\begin{equation}\label{eqn:abcd}
  L_{i,j}=\frac14\big(|a_{ij}+b_{ij}+c_{ij}+d_{ij}|+|a_{ij}-b_{ij}-c_{ij}+d_{ij}|+|a_{ij}+\mi b_{ij}-\mi c_{ij}-d_{ij}|+|a_{ij}-\mi b_{ij}+\mi c_{ij}-d_{ij}|\big)
\end{equation}
where we defined
\begin{equation}
  a_{ij}\coloneqq\alpha_{-+}^i\alpha_{--}^j,\quad
  b_{ij}\coloneqq\alpha_{++}^i\alpha_{+-}^j,\quad
  c_{ij}\coloneqq\alpha_{--}^i\alpha_{-+}^j,\quad
  d_{ij}\coloneqq\alpha_{+-}^i\alpha_{++}^j.
\end{equation}
We give some values of $L_{i,j}$ in \cref{tab:Lij}.

\begin{table}[H]
  \centering
  \begin{tabular}{|c|ccccccccccc|}
    \hline
    \backslashbox{$i\!\!\!\!$}{$\!\!\!\!j$}
       & 0         & 1         & 2         & 3         & 4         & 5         & 6         & 7         & 8         & 9         & 10         \\ \hline
    0  & 1         & 4         & 8         & 24        & 56        & 160       & 384       & 1088      & \bf2624   & 7424      & $17\,920$  \\
    1  & 4         & 4         & 8         & 24        & 64        & 160       & 448       & \bf1088   & 3072      & 7424      & $20\,992$  \\
    2  & 8         & 8         & 8         & 32        & 64        & 192       & \bf448    & 1280      & 3072      & 8704      & $20\,992$  \\
    3  & 24        & 24        & 32        & 32        & 64        & \bf192    & 512       & 1280      & 3584      & 8704      & $24\,576$  \\
    4  & 56        & 64        & 64        & 64        & \bf64     & 256       & 512       & 1536      & 3584      & $10\,240$ & $24\,576$  \\
    5  & 160       & 160       & 192       & \bf192    & 256       & 256       & 512       & 1536      & 4096      & $10\,240$ & $28\,672$  \\
    6  & 384       & 448       & \bf448    & 512       & 512       & 512       & 512       & 2048      & 4096      & $12\,288$ & $28\,672$  \\
    7  & 1088      & \bf1088   & 1280      & 1280      & 1536      & 1536      & 2048      & 2048      & 4096      & $12\,288$ & $32\,768$  \\
    8  & \bf2624   & 3072      & 3072      & 3584      & 3584      & 4096      & 4096      & 4096      & 4096      & $16\,384$ & $32\,768$  \\
    9  & 7424      & 7424      & 8704      & 8704      & $10\,240$ & $10\,240$ & $12\,288$ & $12\,288$ & $16\,384$ & $16\,384$ & $32\,768$  \\
    10 & $17\,920$ & $20\,992$ & $20\,992$ & $24\,576$ & $24\,576$ & $28\,672$ & $28\,672$ & $32\,768$ & $32\,768$ & $32\,768$ & $32\,768$  \\ \hline
  \end{tabular}
  \caption{
    Values of $L_{ij}$ for small $i$ and $j$.
    The antidiagonal in bold corresponds to $N=i+j+1=9$ and reaches its highest value when $i=N-1$ or $j=N-1$, that is, when the deterministic strategies chosen by all parties coincide.
  }
  \label{tab:Lij}
\end{table}

In order to prove the validity of the local bound, we have to show that the pattern emphasised in \cref{tab:Lij} for $N=9$ generalises, that is, that all antidiagonals of this array grow from the diagonal to the edges.
As the form in \cref{eqn:abcd} is not very practical for this purpose, we derive inductive formulas from there.

First, it is clear that
\begin{equation}
  R^2S^2=-8\cdot\id\quad\text{gives}\quad L_{i+2,j+2}=8L_{i,j}.
  \label{eqn:recursive1}
\end{equation}
Second, we can derive the explicit phases involved in the moduli in \cref{eqn:abcd} by manually showing that the pattern has a period of eight (in $i$ or $j$), which in turn allows us, thanks to the newly linearised equation, to derive the following relation:
\begin{equation}
  L_{i,j+4}=8(L_{i,j+2}-L_{i,j}),
  \label{eqn:recursive2}
\end{equation}
which we have to prove separately for odd and even $j$ and for $i=0$ and $i=1$, the other values following from \cref{eqn:recursive1}.
With \cref{eqn:recursive1,eqn:recursive2} at hand, a simple induction suffices to prove the desired result, namely, that the antidiagonals of \cref{tab:Lij} reach their maximum on the edges.

Given the tedium of this proof, it would be interesting for further work to simplify it by connecting it with the abundant literature on generalised Fibonacci numbers and identities involving multinomial coefficients, see~\cite{Er84,WW20}, just to mention a few.
We were unsuccessful in our attempts to do so but we expect elegant methods to easily solve this small case and to generalise to more measurements.
To this end, note that the matrices from \cref{eqn:RS} can be written
\begin{equation}\nonumber
  R=\id+A+A^2+A^3=
  \begin{pmatrix} 1 \\ 1 \\ 1 \\ 1 \end{pmatrix}
  \cdot
  \begin{pmatrix} \id \\ A \\ A^2 \\ A^3 \end{pmatrix}
  \quad\text{and}\quad
  S=\id+A-A^2+A^3=
  \begin{pmatrix} 1 \\ 1 \\ -1 \\ 1 \end{pmatrix}
  \cdot
  \begin{pmatrix} \id \\ A \\ A^2 \\ A^3 \end{pmatrix}
  \quad\text{where}\quad A=
  \begin{pmatrix}
    0 & 1 & 0 & 0 \\
    0 & 0 & 1 & 0 \\
    0 & 0 & 0 & 1 \\
    -1 & 0 & 0 & 0
  \end{pmatrix}.
\end{equation}
The sign pattern is precisely the one of the underlying orbit.
These observations allow us to write, for instance, $R^i$ as a combinatorial sum instead of the analytic expression from \cref{eqn:abcd}, namely,
\begin{equation}
  R^i=\sum_{k_0+k_1+k_2+k_3=i}\binom{i}{k_0,k_1,k_2,k_3}\id^{k_0}A^{k_1}A^{2k_2}A^{3k_3}=\sum_{k_0+k_1+k_2+k_3=i}\frac{i!}{k_0!k_1!k_2!k_3!}A^{k1+2k_2+3k_3},
\end{equation}
which can be simplified further given that $A^4=-\id$.

\section{Full table of results}

\begin{table}[H]
  \centering
  \begin{tabular}{|c|c|c|c|c|c|c|c|c|}
    \hline
    \backslashbox{$m\!\!\!\!\!\!$}{$\!\!\!\!\!\!N$}
           & 3            & 4            & 5            & 6            & 7            & 8            & 9            & 10           \\ \hline
     2     & \bf0.5       & \bf0.5       & \bf0.25      & \bf0.25      & \bf0.125     & \bf0.125     & \bf0.0625    & \bf0.0625    \\
     3     & \bf0.57143   & \bf0.39024   & \bf0.26230   & \bf0.17534   & \bf0.11700   & \bf0.07802   & \bf0.05202   & \bf0.03468   \\
     4     & \bf0.5       & \bf0.35355   & \bf0.21875   & \bf0.15468   & \bf0.09375   & \bf0.06629   & \bf0.04004   & \bf0.02831   \\
     5     & \bf0.50794   & \bf0.34293   & \bf0.22521   & \bf0.14653   & \bf0.09503   & \bf0.06155   & \bf0.03985   & \bf0.02579   \\
     6     & \bf0.49505   & \bf0.33800   & \bf0.21296   & \bf0.14236   & \bf0.08866   & \bf0.05913   & \bf0.03678   & \bf0.02452   \\
     7     & \bf0.49911   & \bf0.33302   & \bf0.21666   & \bf0.13971   & \bf0.08983   & \bf0.05770   & \bf0.03705   & \bf0.02379   \\
     8     & \bf0.49317   & \bf0.33059   & \bf0.21077   & \bf0.13809   & \bf0.08691   & \bf0.05680   & \bf0.03570   & \bf0.02332   \\
     9     & \bf0.49676   & \bf0.32994   & \bf0.21347   & \bf0.13705   & \bf0.08779   & \bf0.05619   & \bf0.03596   & \bf~0.02301~ \\
     10    & \bf0.49206   & \bf0.32836   & \bf0.20985   & \bf0.13623   & \bf0.08612   & \bf0.05575   & \bf~0.03521~ &              \\
     11    & \bf0.49288   & \bf0.32740   & \bf0.21158   & \bf0.13565   & \bf0.08675   & \bf~0.05544~ &              &              \\
     12    & \bf0.49221   & \bf0.32721   & \bf0.20938   & \bf0.13524   & \bf~0.08569~ &              &              &              \\
     13    & \bf0.49249   & \bf0.32637   & \bf0.21062   & \bf0.13488   &              &              &              &              \\
     14    & \bf0.49176   & \bf0.32606   & \bf0.20904   & \bf0.13461   &              &              &              &              \\
     15    & \bf0.49320   & \bf0.32607   & \bf0.21010   & \bf0.13441   &              &              &              &              \\
     16    & \bf0.49160   & \bf0.32548   & \bf0.20884   & \bf~0.13422~ & 0.08526      &              &              &              \\
     17    & \bf0.49192   & \bf0.32530   & \bf0.20961   &              &              &              &              &              \\
     18    & \bf0.49180   & \bf0.32531   & \bf~0.20872~ &              &              &              &              &              \\
     19    & \bf0.49184   & \bf0.32500   & 0.20933      &              &              &              &              &              \\
     20    & \bf0.49153   & \bf0.32493   & 0.20861      &              &              &              &              &              \\
     21    & \bf0.49210   & \bf~0.32494~ & 0.20916      &              &              &              &              &              \\
     22    & \bf~0.49143~ & 0.32471      & 0.20854      &              &              &              &              &              \\
     23    & \bf0.49155   & 0.32463      & 0.20897      &              &              &              &              &              \\
     24    & \bf0.49150   & 0.32465      & 0.20849      & 0.13352      &              &              &              &              \\
     25    & \bf0.49162   &              &              &              &              &              &              &              \\
     26    & \bf0.49143   &              &              &              &              &              &              &              \\
     27    & 0.49162      &              &              &              &              &              &              &              \\
     28    & 0.49141      &              &              &              &              &              &              &              \\
     29    & 0.49147      &              &              &              &              &              &              &              \\
     30    & 0.49145      &              &              &              &              &              &              &              \\
     31    & 0.49146      &              &              &              &              &              &              &              \\
     32    & 0.49138      & 0.32423      & 0.20836      & 0.13326      &              &              &              &              \\
     64    & 0.49134      & 0.32391      & 0.20824      &              &              &              &              &              \\
     96    & 0.49133      & 0.32386      &              &              &              &              &              &              \\
     128   & 0.49133      & 0.32384      &              &              &              &              &              &              \\
     ~224~ & 0.49132      &              &              &              &              &              &              &              \\ \hline
  \end{tabular}
  \caption{
    Critical visibility $v_m^{\GHZ_N}$ for the nonlocality of the $N$-partite GHZ state with each party performing $m$ measurements forming a regular polygon on the XY plane.
    Each value is obtained analytically by finding the facet of the symmetrised local polytope that optimally detects $\br^{(m)}_N$ from \cref{eqn:GHZtensor}.
    The local bound is rigorously established only for the bold values; the other ones rely on a heuristic, although the trivial deterministic strategy always saturates it so that there might be some symmetry arguments enabling us to generalise the proof from \cref{app:m=4}.
  }
  \label{tab:GHZup}
\end{table}

\end{document}